\documentclass[%
reprint, 
 amsmath,amssymb,aps]{revtex4-2}
\usepackage[utf8]{inputenc}
\usepackage{tikz-feynman}
\usepackage{amsfonts}
\usepackage{import}
\usepackage{graphicx}
\usepackage[english]{babel}
\graphicspath{{./drawings/}}

\begin{document}
\preprint{APS/123-QED}
\title{Second Order Twist Contributions to the Balitsky-Kovchegov Equation at small-x: \\ Deterministic and Stochastic Pictures}
\author{S. K. Grossberndt}
\affiliation{Department of Physics, Columbia University}
\date{\today}
\begin{abstract}
    I study the second order twist corrections to a toy model of a dipole-dipole interaction in the context of a both deterministic and stochastic effects. This work is done in the high $N_C$ limit in the Bjorken picture. I show that the correction to the second twist terms of the stochastic picture suggest additional importance of the second twist correction in the stochastic model as compared to the deterministic model. 
\end{abstract}
\maketitle

\section{Introduction}
\hspace{0.5 cm} There has been a large body of work in the field of high energy QCD related to the scattering for virtual photons on bound states \cite{MM} \cite{Mun} \cite{KL} \cite{bj} \cite{kharzeev2002high}. 
Important results in this field include the parton model of Deep Inelastic scattering, first given by Feynman \cite{Feynman} and later modified by Bjorken and Paschos \cite{bjp}, and the DGLAP evolution equations \cite{dglap1}\cite{dglap2}\cite{dglap3}, both cornerstones of perturbative QCD. 
This work was originally performed in Lorentz-covariant Feynman diagrams techniques \cite{St} \cite{PS}, however, one may also proceed with these calculations using light cone perturbation theory as set forth by Lepage and Brodsky \cite{lcpt}, in their paper following the work of Bjorken, Kogut and Soper \cite{bks}.
This method allows the analysis to proceed via the light cone wave function, which allows for the description of the Fock state of a hadron as a function of gluon and quark numbers. 
Further, as outlined in \cite{lcpt}, the use of the light cone gauge, which simplifies the gluon field to two independent components--the transverse components--and a single dependent component. 
Such work often is done using large $N_C$ perturbation theory and the small-x regime \cite{MR},
where the amplitudes may be descried using the Balitsky-Fadin-Kuraev-Lipatov (BFKL) equation \cite{kuraev1977pomeranchuk}\cite{lipatov1976reggeization}.
In these pictures, one must consider the equations via careful choice of approximations and frame.
To this end, one may consider the following simplifications:
take a target that is ultra-relativistic, thus defining the "Infinite Momentum Frame" or Bjorken Frame \cite{bj} and setting the momentum of the target much larger than the mass and the center of mass energy is high (thus the Bjorken-x, which is in exact analogy to the Feynman-x, is small); 
 ignore the evolution of the quark distributions of the hadron in this frame as the distribution functions of such run significantly slower than the gluon distribution; 
 assume that the coupling constant is fixed (or at least $\alpha_s<<1$).
These assumptions give rise to the double logarithmic approximation of the DGLAP evolution equations \cite{KL} .

One may consider DIS in the rest-frame of the nucleon. 
In this frame, the virtual photon fluctuates into a quark-antiquark pair, that, in the large-$N_c$ limit may develop into a cascade of gluons which may be described by the Mueller dipole model \cite{mueller2000small}. 
When resumation is performed on this cascade, one reaches the Balitsky-Kovchegov (BK) evolution equation which is unitary and does not diffuse into the IR,  distinguishing it from earlier equations and generates a saturation scale that grows with energy: allowing for the suppression of the non-perturbative QCD physics that may enter through the initial conditions  \cite{kovchegov2000unitarization}\cite{balitsky1996operator} \cite{Golec-Biernat2004} \cite{Levin2005}.
As described by Balitsky, the relevant logarithms in this equation may arise from the expansion of non-local operators which gives rise to a twist interpretation of the equation, where higher twists represent the original dipole branching and further scattering off of the nucleon \cite{balitsky1996operator} \cite{Levin2005}. 
Balitsky developed the structure functions as a sum over integer moments in the factorization theorem as  a twist expansion, with the twists corresponding to damping by powers of $\frac{1}{Q^2}$, an approach refined and expanded upon in \cite{Kutak2012} \cite{Harris2020} \cite{Kumano2020} \cite{Bastami2020} .
This twist expansion appears explicitly in the BFKL approach, the anomalous dimension of the twist-2 operators sets the $Q^2$ dependence in the deep inelastic moments \cite{Ioffea}. 
In the derivation of equation 2 of \cite{balitsky1996operator}, the use of light cone perturbation theory allows for the factorization of the relevant diagrams \cite{kovchegov2000unitarization} which often allows for the diagrams to be represented in the form of Muller Vertices \cite{St}.

However, this equation assumes a linear scaling in the bulk of the phase space, which allows for unrestrained growth of density of the dipoles in this equation. 
To correct for this, one argues for a modeling of the non-linearity using recombination in analogy to branching-diffusion models. 
This process is captured using stochastic corrections of tip fluctuations, which may not be modeled analytically, and front fluctuations, which will be considered in this paper \cite{Mun}. There has been recent work in Monte Carlo analysis to correspond to analytical calculations at infinite-time limits \cite{Le2020}.
Such corrections are motivated by their applicability to the Fischer-Kolmogorov-Petrivsky-Piscounouv equation and its mapping to the BK equation \cite{Wang2020}. 
Thus, for the purposes of this paper, I shall consider the equations without such corrections to be the "Deterministic equations" and those with such to be the "Stochastic equations". 

This paper aims to connect the work done on these fluctuations on the second order twist corrections with the operator language of Balitsky \cite{balitsky1996operator}. This will provide corrections to the DGLAP equation to account for lower values of $Q^2$, taking the BFKL approach rather than via string operators, which are an equivalent approach to higher twist correction to DIS \cite{Chirilli2010}.  
This paper takes an approach similar to \cite{Motyka2014}, further connecting the OPE and factorization approaches, then extending into the stochastic world.
\section{Calculating the Second Order Twist}
\hspace{0.5 cm} To investigate the dynamics of the second order twist contribution in the stochastic picture, I must first evaluate the deterministic picture.
In order to do this, I utilize a simplified picture where the probe may be one or more dipoles prepared in some manner that is irrelevant to this discussion and the scatting target is an individual dipole. 
This toy model will allow us to consider the relevant evolution equations without needing to concern ourselves with the physical compilations of true Deep Inelastic Scattering or the process by which an additional dipole is produced from the original. 
Then, define the contribution from each twist to the forward scattering amplitude, defined as 
\begin{equation}
    T= 1-S \hspace{2 cm} T=T_d+T_{dd}+T_{ddd}+....
\end{equation}
With $T_d$ being the contribution from a single dipole (leading twist), $T_{dd}$ from two dipoles (higher twist) and so forth. 
The first twist scattering amplitude of a dipole with transverse size $x_{\perp}$ and Bjorken variable x is given as \cite{FD}. \\
\begin{figure}[h!]
    \centering

\includegraphics{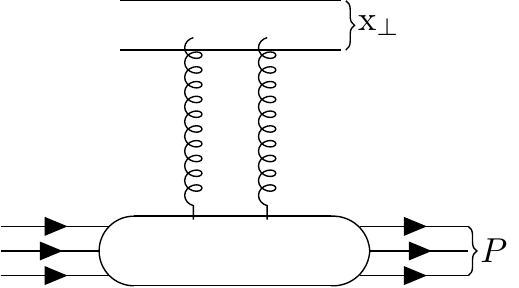}
 \begin{eqnarray}
  T_d(x, \frac{1}{x_\perp^2})= \frac{1}{2 \pi P^+} \int d \xi_- e^{-i P^+ x \xi_- } \nonumber \\ \times \left< P | F^a_{+ \alpha} (0, \xi_-, x_{\perp}) \gamma^+ W_{ab} F^b_{+\alpha}(0)|P \right> 
	 \label{eqn:tmd1}
\end{eqnarray}
    \caption{Lowest Twist Gluon Distribution. }
    \label{fig:logd}
\end{figure}
Which gives 
\begin{equation}
    T_d=\frac{\alpha_s \pi^2}{2 N_C}x_{\perp}^2 xG(x, \frac{1}{x_{\perp}}^2)
	\label{eqn:into}
\end{equation}
I require the gluons to form a color singlet, as the dipole must also be a color singlet. 
By setting the scattering off of a dipole, rather than utilizing a Muller vertex, I escape the divergences that necessitate renormalization at scale $Q$, but this scale is identified with $\frac{1}{x^2_{\perp}}= Q^2$ as in \cite{KL}. W is the "gauge link" or Wilson operator, which by choice of the gauge $A_+=0$, $\gamma^+ W_{ab}$ may be set to 1. 
The choice of this gauge is the light cone gauge used in \cite{KL} and is chosen as to connect equation 2 to equation 3. 
This connection is motivated in \cite{FD}, and uses the fact that the light cone gauge gives a representation of hadronic states in terms of the partonic basis that does set the Weizs{\"a}cker-Williams distribution as the number density of gluons.
Equation \ref{eqn:tmd1} is the transform of a Transverse Momentum Dependent distribution which is not gauge invariant when utilized to count the number of gluons; a correction for which involves choosing the gauge link to cancel in the particular gauge, thus, leading to the method outlined in \cite{FD}, and is seen to be equivalent to the Wilson line approach in \cite{Balitsky2017}.
This process is illustrated in figure \ref{fig:logd}, with the relevant integral to be calculated.
In this gauge, 
$$
F_{+ \alpha}= \partial_+ A_{\alpha}
$$
Using this approach allows for the application of Operator Product Expansion to derive equation \ref{eqn:into} from equation \ref{eqn:tmd1}.  
For the next-order twist, I consider the case of a state of two dipoles scattering off of the target proton. 
For the purposes of this discussion, I shall consider the method of preparation of this system to be irrelevant, 
but I require that the impact parameter of both dipoles be equal to zero, and that the dipoles have transverse size $x_{1 \perp}$ and $x_{2 \perp}$ respectively.
This results in a system where a set of two dipoles each scatter off of the target once as illustrated in figure \ref{fig:sotd}.
This yields the factorized equation,
\begin{figure}[h!]
    \centering
    \begin{tikzpicture}
\begin{feynman}
\vertex(a);
\vertex[below right =1 cm of a](b);
\vertex[below=3 cm of a] (c);
\vertex[above=0.25 cm of c](d);
\vertex[below=0.25 cm of c](e);
\vertex[right=1 cm of c](f);
\vertex[above right =0.25 cm and 0.1 cm of f](g);
\vertex[below right=0.25 cm and 0.1 of f](h);
\vertex[above right= 0.05 cm and 0.5 cm of b](i);
\vertex[below right= 0.05 cm and 0.5 cm of b](i1);
\vertex[above right= 0.5 cm of b](j);
\vertex[below right=0.5cm of b](k);
\vertex[right= 2.25 cm of j](l);
\vertex[right=2.25 cm of k](m);
\vertex[right=2 cm of i](n);
\vertex[right= 2. cm of i1](n1);
\vertex[right= 2.75 cm of b](o);
\vertex[right=4.25 of a](p);
\vertex[above right=0.5 cm of f](r);
\vertex[below right=0.5 cm of f](s);
\vertex[right=2cm of r](t);
\vertex[right=2cm of s](u);
\vertex[right=2.75 cm of f](v);
\vertex[above left =0.25 cm and 0.1 cm of v](w);
\vertex[below left=0.25 cm and 0.1 of v](x);
\vertex[right=1 cm of v](y);
\vertex[above=0.25 cm of y](z);
\vertex[below=0.25 cm of y](a1);
\vertex[above right=0.125 and 0.25 cm of k](g1); 
\vertex[below= 1.85 cm of g1](g2);
\vertex[right= 0.5 cm of g1](g3);
\vertex[above right= 0.45 cm and 0.75 cm of g3](g4);
\vertex[right= 0.5 cm of g4](g5);
\vertex[right= 0.5 cm of g2](g6);
\vertex[right= 0.75 cm of g6](g7);
\vertex[right= 0.5 cm of g7](g8);
\vertex[below=0.4 cm of l](n2);
\vertex[below=0.05 cm of n2](n3);
\diagram*{
(j)--(l);
(k)--(m);
(i)--(n);
(i1)--(n1);
(c)--[fermion](f);
(d)--[fermion](g);
(e)--[fermion](h);
(f)--[quarter left](r)--(t)--[quarter left](v);
(f)--[quarter right](s)--(u)--[quarter right](v);
(v)--[fermion](y);
(w)--[fermion](z);
(x)--[fermion](a1);
(g1)--[gluon](g2);
(g3)--[gluon](g6);
(g4)--[gluon](g7);
(g5)--[gluon](g8);
};
\draw[decoration={brace}, decorate](z.north west)--(a1.south west)
    node[pos=0.5, right]{\(P\)};
\draw[decoration={brace}, decorate](l.north west)--(n2.south west)
    node[pos=0.5, right]{ \(x_{1 \perp}\)};
    \draw[decoration={brace}, decorate](n3.north west)--(m.south west)
    node[pos=0.5, right]{\(x_{2\perp}\)};
\end{feynman}
\end{tikzpicture}
 \begin{eqnarray}
               T_{dd}\left(x, x_1, \frac{1}{x_{1\perp}^2}, \frac{1}{x_{2\perp}^2}\right)= \frac{1}{p^+} \int d \xi d \xi^1 e^{i p^+ x \xi_- } e^{i p^+ x_1 \xi^1_- }\nonumber \\ \times \left< P | F^a_{+\alpha} ( 0, \xi_-, x_{1 \perp})F^a_{+\alpha} (0)F^b_{+\beta} ( 0, \xi^1_-, x_{2 \perp} ) F^b_{+\beta}(0)|P \right> 
 \end{eqnarray}

     \caption{Second order Twist }
     \label{fig:sotd}
 \end{figure}
 //
This yields in the mean field picture, following similar analysis to that for the quark distribution in \cite{St}, and noting a $C_F$ arises from the splitting of the state into two dipoles as can be seen in equation (74) of \cite{Mun},  
\begin{equation}
    \begin{array}{c}
    T_{dd}=  C_F T_d(x,\frac{1}{x_{1\perp}^2}) T_d(x_1,\frac{1}{x_{2\perp}^2}) 
    \end{array}
\end{equation}

Which, taking $x_1 \approx x$ and $x_{1\perp}\approx x_{2\perp}\approx \frac{1}{Q^2}$
\begin{equation}
T_{dd}= \frac{\alpha_s^2 C_F \pi^4}{4 N_C^2} x_{1 \perp}^2 x_{2 \perp}^2 xG^{2}(x, Q^2)
\end{equation}

One notes that by starting with the dipoles independent of the method of preparation of the double dipole system, one suppresses the anomalous dimension arising in the Operator Product Expansion.
However, this is irrelevant for this discussion, as the discussion focuses on simply calculating the interaction of a double dipole interaction to analyze stochastic effects, which are independent of the quadratic terms. 

\section{Deterministic Equations in the Scaling Region}
The twist expansion becomes important in the scaling region, i.e. where the dipole sign is smaller that the inverse saturation momentum. 
In this region, the scattering amplitudes corresponds to the leading twist, with the higher twist terms simply acting as a correction. 

Then, in \cite{KL} the equation for T in the extended geometric scaling region ($x_{\perp} \lesssim 1/Q_{S0}$) becomes 
\begin{equation}
	T_d= (x_{\perp}Q_{s0})^{1+2 i \nu_{sp}} C(\alpha_s) e^{\bar{\alpha}_S y \chi(0, \nu_{sp})}
\end{equation}
Where y is the rapidity and is equal to $ln(x^2E^2)$, where E is the center of mass energy for the dipoles, $\bar{\alpha}_S= \frac{\alpha_S N_C}{\pi}$ and $\chi(0, \nu_{sp})$ is an eigenvalue of the BK Kernel.
Using \cite{Mun} to further refine this equation as is done in equation (118), and noting that the logarithmic term comes from the integral arising in the BK equation as in equation (120),
I then write, with $\gamma_0= \frac{1}{2}+i \nu_{sp}$, and $C(\alpha_s)$ being some constant of the form $\alpha_S \times K$ 

\begin{eqnarray}
    T_d(x_{\perp}, y)=C(\alpha_s) \ln( x_{\perp}^2 Q_S(y)^2) (x Q)^{2\gamma} \nonumber \\ \times \hspace {0.2 cm}  e^{-\frac{\ln^2( x_{\perp}^2 Q_S(y)^2) }{2 \bar{\alpha}_S \chi''(\gamma_0) y}} 
\end{eqnarray} 

Which, substituting for $Q_S(y)=Q_{s0} e^{\bar{\alpha}_S \frac{\chi(\gamma_0)}{2 \gamma_0}y}$ yields
\begin{eqnarray}
	T_d(x_{\perp}, y)=C(\alpha_s) \ln(\frac{1} {x_{\perp}^2 Q_{s0}^2})(x_{\perp}Q_{s0})^{2\gamma} \nonumber \\ 
	\times \hspace{0.2 cm} e^{-\frac{\left(\gamma_0 \ln(\frac{1}{x_{\perp}^2 Q_{S0}^2})+\bar{\alpha}_S \chi(\gamma_0)y\right)^2}{2 \gamma_0 \bar{\alpha}_S \chi''(\gamma_0) y}} 
\end{eqnarray}
\begin{figure}
    \centering
    \includegraphics[scale=0.45]{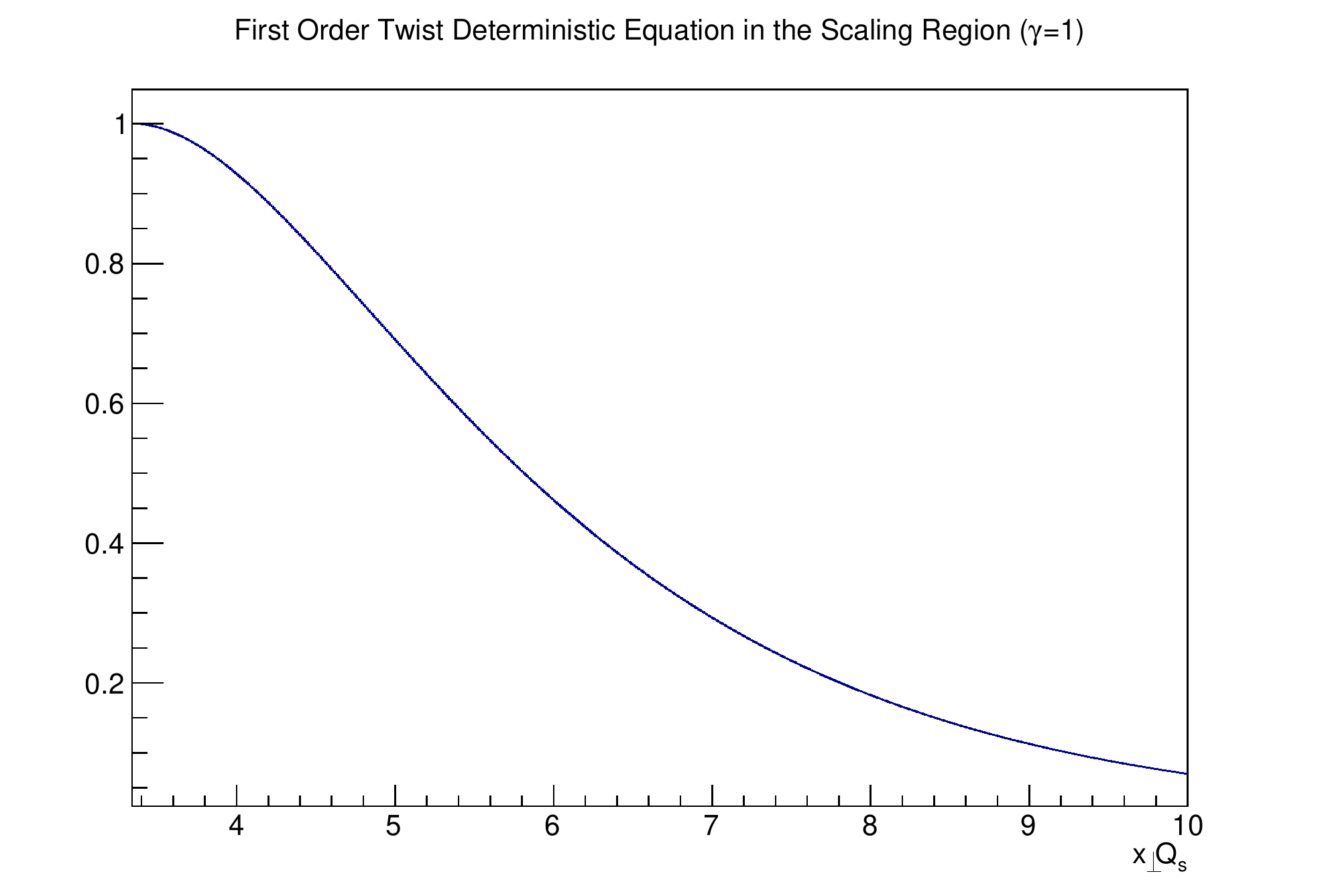}
    \caption{First Order Twist Deterministic Equation. Equation had been normalized to 1 and only relevant geometric scaling region is shown.}
    \label{fig:fotde}
\end{figure}
One notes that, for $x_{\perp}Q_S \lesssim 1$ the final term is very close to 1, thus I shall absorb it into the C term from this point forward. 
Figure \ref{fig:fotde} shows a representation of equation 9 with normalization set such that the max value is 1 with $\gamma$ set to 1 for simplicity of presentation. 
One notes that this representation additionally simplified the final term of equation 9, the leading term of which gives the characteristic shape in the scaling region. 
Thus, by applying the above to equation 6, the higher twist term, following the earlier approach reads off as (with the $C^2$ absorbing all leading terms)
\begin{eqnarray}
	T_{dd}=C^2 \ln^2( x_{\perp}^2 Q_{s0}(y)^2)[x_{1 \perp} x_{2 \perp} Q_{s0}^2]^{2 \gamma} 
\end{eqnarray}
\begin{figure}[h!]
    \centering
    \includegraphics[scale=0.45]{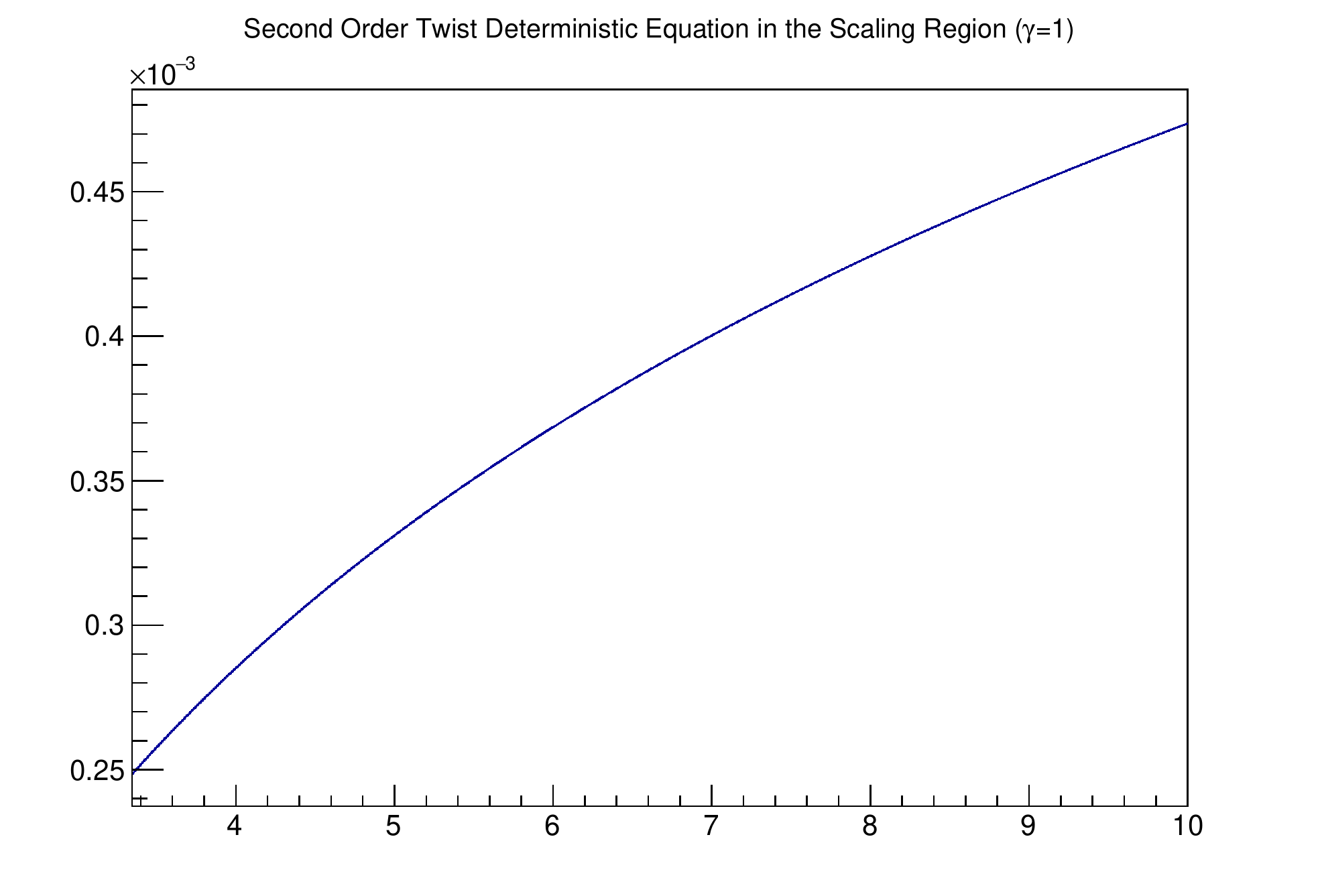}
    \caption{Second Order Twist Deterministic Equation. Normalization is consistent with that applied to figure \ref{fig:fotde} above}
    \label{fig:sotde}
\end{figure}
Figure \ref{fig:sotde} is another simplified representation of the deterministic equations, this time the second order twist in equation 10, which is scaled using the same normalization as figure \ref{fig:fotde}. 
One can see that the corrections from the second order twist are a factor of $10^3$ smaller compared to the first order twist. 
This gives a relative correction, with taking the final term of equation 9 to be $\approx 1$ and taking $x_{1 \perp} \approx x_{2 \perp} \approx x_{\perp}$, of 
\begin{eqnarray}
	T_{dd}/T_{d} \propto (x_{\perp} Q_{s0})^{2 \gamma} \ln \left(x_{\perp} Q_{s0} \right)
\end{eqnarray}
\section{Stochastic effects in the scaling region}
From Munier \cite{Mun}, the equation for scattering amplitude accounting for front fluctuations ($T^S_d$ stands for the Stochastic version of $T_d$ which will be noted as $T^D_d$ for clarity henceforth), with 
$P(\delta)= e^{-\gamma\delta}$ is the probability of having a front delayed by $\delta$ \cite{Mun}.
\begin{eqnarray}
	T_d^S \propto \int_0^{\ln\left(\frac{1}{x_\perp^2Q_{S0}^2}\right)} d \delta P(\delta) \left[\ln( x_{perp}^2 Q_S(y)^2)-\delta\right] \nonumber\\  
     \times \hspace{0.1 cm} e^{-\gamma[\ln(\frac{1}{x^2 Q^2}-\delta]} 
\\
	\propto  [x_{\perp} Q_{S0}^2]^{2 \gamma}  \int_0^{\ln \left( \frac{1}{x_{\perp}^2 Q_S(y)^2}\right)} d \delta \ln(\frac{1} {x_{\perp}^2 Q_S(y)^2})- \delta \\
	\propto \frac{1}{2} \ln^2( \frac{1}{x_{\perp}^2 Q_S(y)^2})[x_{\perp} Q_{s0}]^{2 \gamma}
\end{eqnarray}
Thus 
\begin{equation}
	T_d^S/T_d^D \propto \ln(\frac{1}{x_\perp^2 Q_{S0}^2})
\end{equation}
\begin{figure}
    \centering
    \includegraphics[scale=0.45]{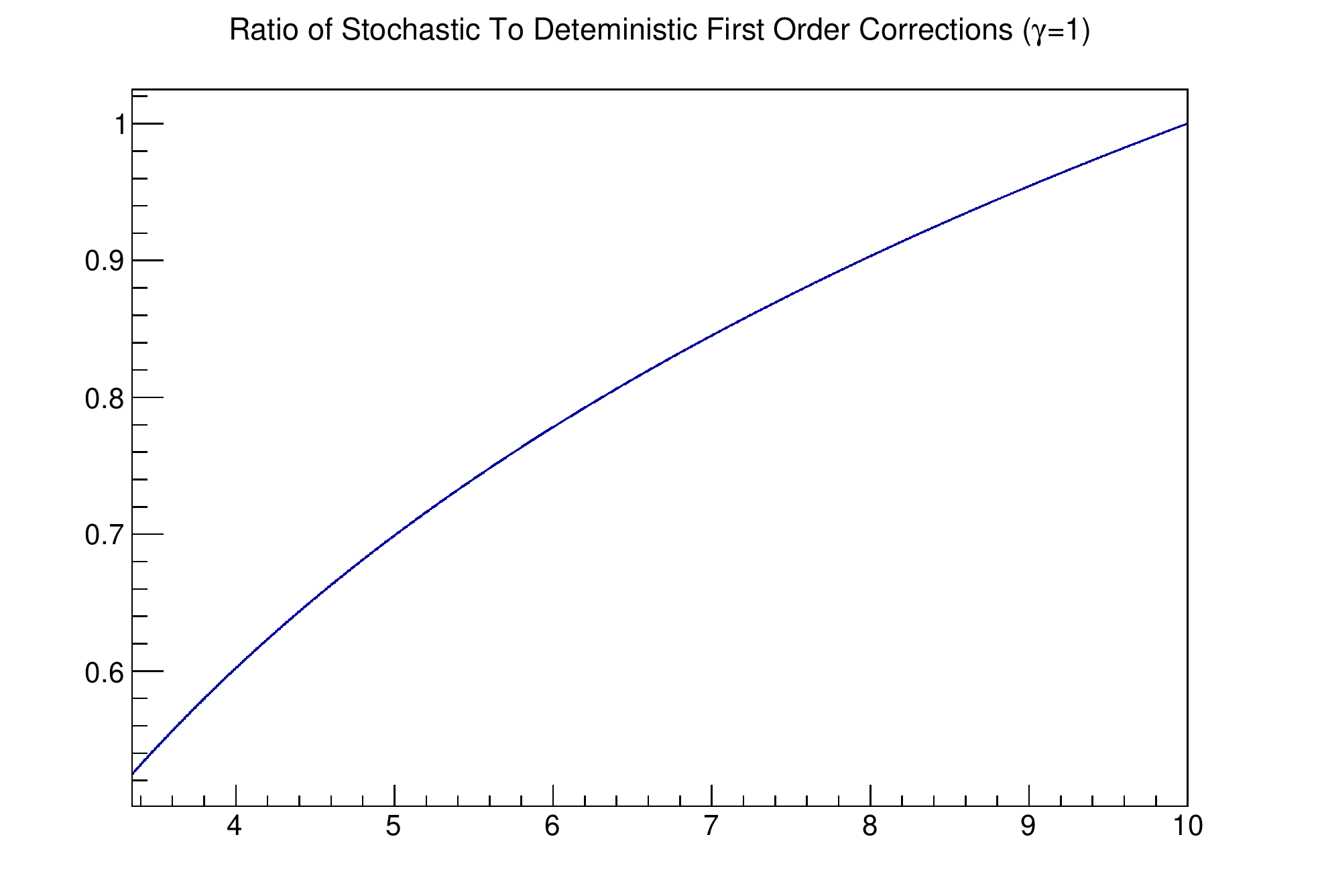}
    \caption{Ratio of Stochastic to Deterministic Equations in the First Order Twist}
    \label{fig:rat1}
\end{figure}
Then, in the second order twist, writing $\ln(\frac{1}{x_{\perp}^2Q_{s0}^2})= \phi \approx \ln(\frac{1}{x_{1 \perp}^2Q_{s0}^2}) \approx \ln(\frac{1}{x_{2\perp}^2Q_{s0}^2})$
\begin{equation}
    \begin{array}{c}
	    T_{dd}^S= \int_0^\phi d \delta e^{-\gamma \delta} C^2 \left[\ln( \frac{1}{x_{\perp}^2 Q_S(y)^2})-\delta\right]^2 e^{-2\gamma [\ln(\frac{1}{x_{1\perp} x_{2 \perp} Q_{s0}^2})- \delta]}\\
\end{array}
\end{equation}

Which taking $\delta \approx \phi+ \textsc{Constant}$, which is to say large fluctuations which dominate other contributions, thus giving an approximate equation of 
\begin{equation}
	T_{dd}^S \approx C^2e^{-2}(\frac{2}{\gamma})^2 (x_{\perp}^2 Q_{S0}^2)^{ 2\gamma}
\end{equation}
Thus 
\begin{equation}
    T_{dd}^S/T_{dd}^D \propto \frac{1}{\gamma (x_{\perp}Q_S)^{2\gamma}\ln( x_{\perp}^2Q_S^2)}
    \end{equation}
\begin{figure}
    \centering
    \includegraphics[scale=0.45]{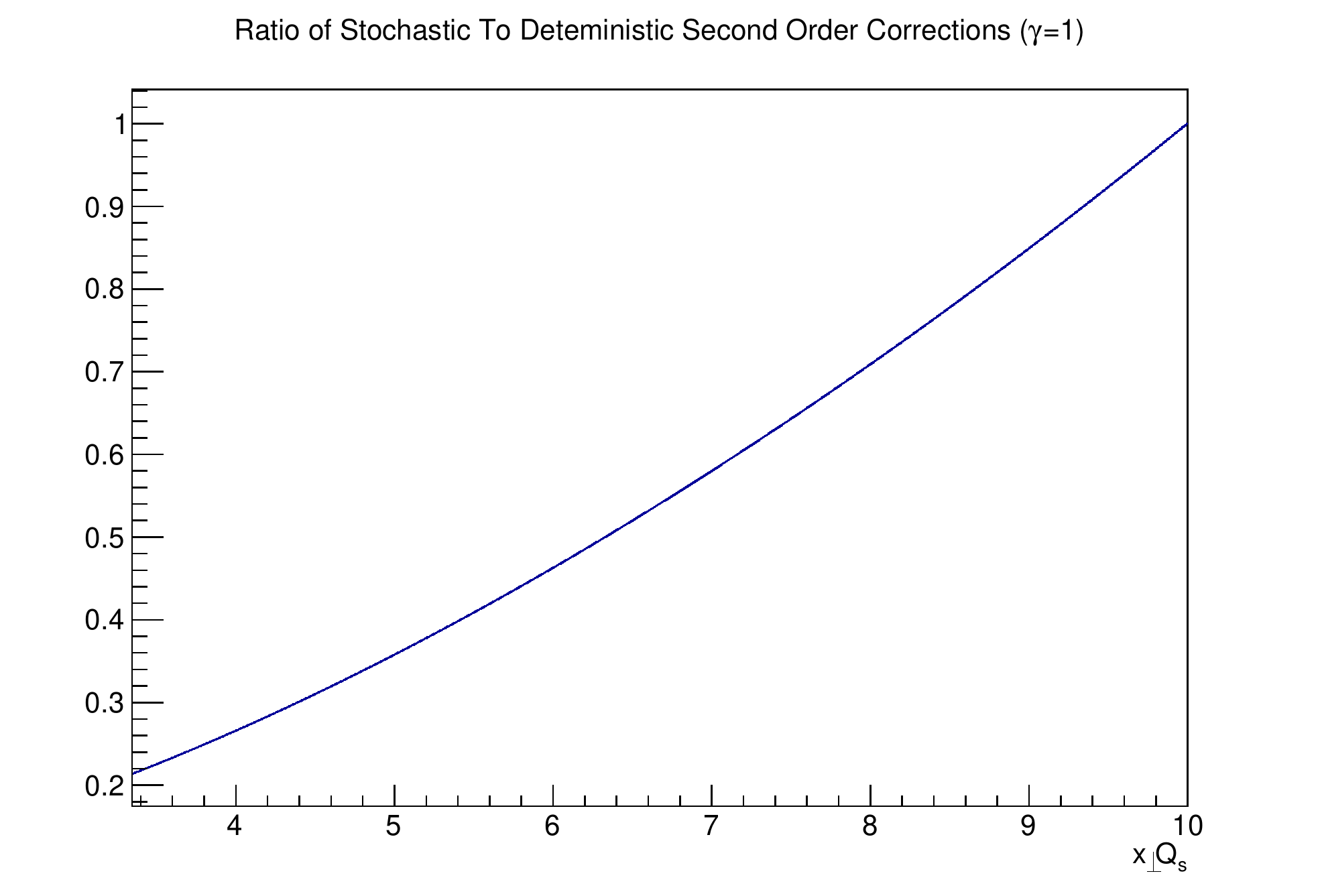}
    \caption{Ratio of Stochastic to Deterministic Equations in the Second Order Twist}
    \label{fig:rat2}
\end{figure}
    \begin{equation}
        T_{dd}^S/T_d^S \propto \frac{1}{\ln\left(\frac{1}{x_{\perp}^2 Q_S^2}\right)}
\end{equation}
    \begin{figure}
        \centering
        \includegraphics[scale=0.45]{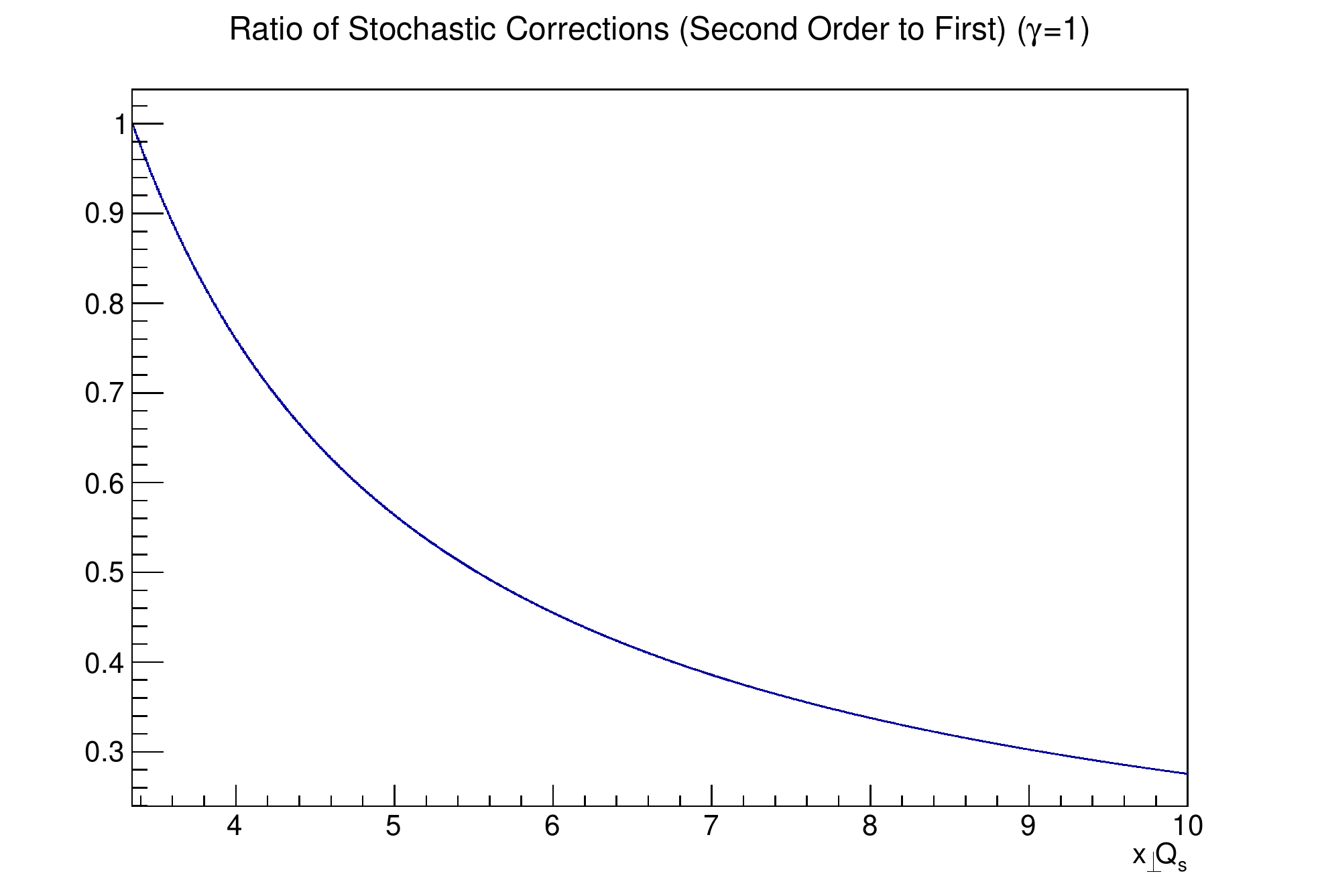}
        \caption{Ratio of Stochastic Equations in the Scaling Region}
        \label{fig:stochrat}
    \end{figure}

Figures \ref{fig:rat1}-\ref{fig:stochrat} represent ratios of stochastic equations, with the max of each ratio in the relevant ranges--keeping with those set in figure \ref{fig:sotde}--set to 1. Figure \ref{fig:rat1} is the ratios of the first order twist equations, and figure \ref{fig:rat2} is the same for the second order twist. 
One can see that the ratios of terms grows substantially more quickly in the second order twist. 
As expected, the stochastic terms contribute larger corrections as xQ grows. 
Figure \ref{fig:stochrat} represents the ratios of stochastic corrections (Second order divided by first order). 

The most interesting result of this analysis is that the second order stochastic term is larger than the factorization approach correction given by $(T_d^S)^2$ by a factor of 
\begin{equation}
	\left(\frac{1}{\gamma \ln^2( x_{\perp}^2 Q_{s0} )}\right)^2 (T_{d}^S)^2 = T_{dd}^S 
\end{equation}
This then gives that there is a correction about equivalent to the square of the gluon distribution function, implying that the dynamics of the scattering may need corrections corresponding to ladder diagrams, thus connecting to approaches in BFKL and GGM \cite{KL}.
\section{Summary and Conclusions}
\hspace{0.5 cm} By implementing the stochastic corrections to the second-order twist of a dipole-dipole interaction, it is apparent that, in the stochastic region, the factorization approach does not properly capture the dynamics, leading to a large correction on the second order term. 
Further studies in this area, including fixing leading terms in a precise DIS context would allow for a better understanding of this effect. 
This work has translated the work previously done on the fluctuations of higher twists to the operator representation of the gluon distribution, as is seen in \cite{Balitsky2017} \cite{duclou2017deep} \cite{duclou2017deep}. 
Further, this work shows, through this connection, that the corrections to the scattering amplitude from the second order twist stochastic component is inversely proportional to the gluon distribution of the target dipole. 

In this work, I have presented a proof-of-concept of an approach that may be applied to later studies to solve the problem of a Stochastic version of the BK evolution equations. 
Additionally this work may be implemented in the study of proper Deep-Inelastic scattering rather than in the toy model presented in this paper. 

Further work in this vein would include a mechanism by which to measure this effect as including effects of nuclear fluctuations on small nuclei. 
Such work would require similar analysis of the JIMWLK equation.
\section{Acknowledgments}
I would like to thank Alfred Mueller of Columbia University for his invaluable help with this paper. Without his support this paper could not have been written.
\bibliographystyle{ieeetr}
\bibliography{main}
\end{document}